\documentstyle[12pt,a4]{article}
\input{epsf.tex}
\begin{document}
\begin{center}
{\bf Some problems of extracting level density and radiative strength
functions from the $\gamma$-spectra in nuclear reactions
}
\end{center}
\begin{center}
{V.A. Khitrov, A.M. Sukhovoj}\\
{\it Joint Institute for Nuclear Research, Dubna, Russia}\\
{Pham Dinh Khang, Vo Thi Anh}\\
{ \it Hanoi National University}\\
{ Vuong Huu Tan, Nguyen Canh Hai, Nguyen Xuan Hai
}\\
{\it Vietnam Atomic Energy Commission}\\
\end{center}
\hspace*{16pt}

\bigskip
\parbox{155mm}
{The most important systematical errors in determination of level density
and radiative strength functions for deformed nuclei have been estimated
from the  gamma-ray spectra in nuclear reactions like the stripping or
pickup reactions.}

\section{Introduction}
At the excitations below the neutron binding energy $E_{ex}\leq B_n$ the
radical complication of the structure of the nuclear excited states occurs.
Simple, practically one-component low-lying levels transform into the
extremely complicated Bohr's high-lying compound states. For understanding
of the main properties of this transition it is necessary to determine
density $\rho$ of levels for a given interval of their quantum numbers
and the mean widths of populating and depopulating them
$\gamma$-transitions, at least, with systematical error provided by
the modern experimental technique. Only on the basis of such
reliable and precise information one can develop adequate theoretical
model of the process under consideration. Unfortunately, more or less
systematical errors are inherency of any method for data treatment.

The reliable and precise data on level density $\rho$ and probability of emission
of any reaction products are important not only for verification and
development of theoretical notions but also from the practical point of view.
In particular, reliable and precise enough evaluation of interaction
cross-section of neutrons with fissile nuclei will permit one to raise the
efficiency of exploitation of nuclear reactors.
Modern analysis [1] of a bulk of the
experimental data show, however, that the quality of the evaluation of the
required parameters depends on the degree of correspondence of the used model
notions to the experiment.

The analysis [2] of these cross-sections with a great probability points out at
rather considerable underestimate of the  pairing interaction effect by
existing models and, as a consequence, at visible discrepancy of conventional
notions of the energy dependence of level density, of at least, fissile
nuclei with the experiment.

The problem of reliable experimental determination of the level density
excited in some nuclear reaction below the energy $E_{ex} \simeq 6-8$ MeV
is not solved yet. The analysis of evaporation spectra in reactions
like $(n,n^{'})$, $(p,n)$ and so on used for this aim contains unknown
systematical errors caused by impossibility of correct accounting for the
reaction mechanism and precise calculation of the penetration coefficients of
nuclear surface for reaction products below $B_n$. As a result, corresponding
experimental data can contain strongly underestimated inherent uncertainty.
This problem is, probably, of special importance [3] for low (3-4 MeV)
excitation energy of a nucleus - reaction product. This quite unambiguously
follows from the fact that the areas of spectra in method [3] are (in the
first approach) inversely proportional to the level density, but just in this
excitation region of deformed nuclei this method has maximal sensitivity,
and the maximal discrepancy between the obtained $\rho$ is observed.

The systematical errors manifest themselves in any other experiments, for example,
in investigations of intensity $I_{\gamma \gamma}$ of two-step cascades [3],
and in analysis of spectra $S$ of $\gamma$-transitions depopulating levels
with energies $E_i$ excited in the (d,p) [4] or $(^3He,\alpha)$ reactions
[5,6].

The last method must give also considerable systematical errors in
determination of the absolute width of $\gamma$-transitions with the
energy $E_{\gamma}\leq E_i$ owing to their strong correlations with errors
of level density (and vice versa).

\section{On some systematical errors in determination of $\rho$ and $k$
in modern experiment}

It is obvious that the correspondence between  $\rho$ and radiative strength
function
$k=f/A^{2/3}=\Gamma_{\lambda}/(D_{\lambda} E^3_{\gamma}A^{2/3})$
 obtained within
different methods in the limits of corresponding uncertainties is one of
criterions of their confidence. Unfortunately, the $\rho$ and $k$ values
determined according [3] and [5,6] do not obey this condition.
So, the $I_{\gamma\gamma}$ values cannot be reproduced within experimental
accuracy using level density from [5,6] due to strong difference
with $\rho$ from [3]. For example, the main peculiarity of the data [2,3] ---
clearly expressed ``step-like" structure in $\rho$ --- practically
 is absent in the data [5,6]. Comprehensive enough analysis of systematical
errors of $\rho$ and $k$ determined with the use of the two-step cascade
intensities  was performed in [3]. But it cannot explain the source of this
discrepancy of the observed results.

One of possible explanations for discrepancy between the results [5,6] and
[3] can be obtained from analysis of the suggested in Oslo method of
extraction of the desired parameters which provide minimum value for $\chi^2$:
\begin{equation}
\frac{\chi^2}{N}=\frac {1}{N} \sum_{E_i} \sum_{E_{\gamma}}
(\frac {S_{th}(E_i,E_{\gamma})-S_{exp}(E_i,E_{\gamma})}{\Delta S(E_i,E_{\gamma})})^2
\end{equation}
for the system of equations connecting experimental $\gamma$-spectra $S$
with the desired parameters $\rho$ and $k$ marked below as $X$:
\begin{equation}
S(X)=S_{th}(E_i,E_{\gamma})=\Gamma_i \times \rho_i/\sum(\Gamma_i \times \rho_i)
\end{equation}
As it was pointed out in [4], this system can be complemented by the
expression for the experimental value of the total radiative width of
neutron  resonance. It should be noted that the system of any  non-linear
equations  (including system (1))  is solved [7] in the framework of
matrix algebra using the Gauss modified (regularized) method and does not
require any additional algorithms for fitting (like that suggested in
[5]).

The value of vector-column $X$ consisting from $n$ parameters is determined [7]
in the vicinity of their actual values for $k+1$ iteration by the matrix
equation
\begin{equation}
X_{k+1}=X_{k}-(J^{T}GJ)^{-1} J^{T} G S(X_{k}),
\end{equation}
where $G$ is the matrix of weights, the Jacobi matrix $J$ and corresponding
transposed matrix  $J^{T}$ are the matrixes of derivatives from function
(2)  with respect to the desired radiative strength function $k$ of
transition and number $\rho$ of levels in a given energy interval of
corresponding spectrum. $S$ is the vector-row of $m$ experimental points in
all spectra used for fitting of parameters. It is obvious [7] that eq.~(1)
has a solution (unique) only upon condition of existence of covariant matrix
$C=(J^{T}GJ)^{-1}$.

Derivatives with respect to $n$ desired parameters in $m$ ($m \ge n$)
points are determined analytically from eq.~(2). Products of  $\chi^2/N$ on
diagonal elements of matrix
$C$ in point of minimum of (1) equal [7] variances of desired parameters,
on non-diagonal elements equal covariants of these parameters. It is well
known in mathematical statistics [8] that the shape of the multi-dimensional
Gauss distribution used in the general case as a likelihood function depends
also on non-diagonal elements of covariant matrix. In particular, eq.~(1)
has minimum only if covariant matrix slightly differs from diagonal one.
This is clearly seen from eq.~(4) for the Gauss distribution of two variables
$u$ and $v$, which equal random deviations of random values $X$ from the
mathematical expectation $<X>$:
\begin{equation}
F(u,v)=exp(-0.5 ((u^2/\sigma_u^2-2 r uv/(\sigma_v \sigma_v)+y^2/\sigma_v^2))/
(1-r^2))/(2 \pi \sigma_v \sigma_v (1-r^2))
\end{equation}
If correlation coefficient equals unity $r = \pm 1$, then covariant matrix
becomes singular, and maximum of function $F(u,v)$  can be obtained for any values
of $u$ and $v$ which obey equation 
$u/\sigma_u \pm v/ \sigma_v=0$.
 At $r=0$ it exists, of
course, only for $u=0$ and $v=0$.

Correspondingly, at $r=1$ the same maximum of the function is observed for a
multitude values of variables. 
This situation, of course, is
taken into account in any modern program of multi-parameter fitting. It
should be noted that inversion of large matrix at $|r|\approx 1$ also brings
to considerable loss in accuracy.

Conditions for invertibility of matrix $(J^{T}GJ)$ for $^{172}Yb$
considered  in [5] were studied by us for
the model calculated spectra (2) with variations of possible accuracy in
determination of the $S$ values. It was established that even at 0.01\%
relative uncertainty of each point in the spectrum, the matrix $(J^{T}GJ)$
is singular in the limits of precision of modern computers.

The corresponding covariant matrix includes a large enough number of correlation
coefficients which are near or equal to unity (in limits of
calculation precision)
 and therefore it is singular. According to definition [8] of
covariant matrix there is a multitude of ensembles of different values  of
parameters $\rho$ and $k$ which lead to the same magnitude of  $\chi^2$
in (1).

The complete compensation of random deviation of one parameter from its
experimental value by corresponding deviation of other parameter is
possible in the only case when $|r|= 1$. However, change in the desired
parameters leads to the change in the correlation coefficients due to
nonlinear connection between level density and radiative strength functions.
As a consequence, the region of possible change in level density completely
compensated by change in strength functions is limited by some intervals of
values of $\rho$ and $k$ for a given excitation energy of a nucleus.
But the correlation coefficients approach on absolute value to unity at other
values of desired parameters from eq.~(2), and covariant matrix stay
singular.

Therefore, at the analysis of the $S_{exp}$ values it should be taken into
account the possibility of existence of the region of possible values $\rho$
and $k$  where is observed complete mutual compensation of their random
deviations from true value.

A simple possibility to test this statement is provided by the Monte-Carlo
method. Some variant of this method was used by authors [3] namely fir
determination of region of possible $\rho$ and $k$ values which allow
reproduction of the same value of $I_{\gamma \gamma}$.
 Fig.~1 shows several pairs of random functional dependencies
$\rho$ and $k$ obtained in accordance with the iterative procedure [3] for
different input values of these parameters. The spectra $S_{if}$ (2) is
calculated using these parameters are represented in Fig.~2. It is natural
that all pairs of $\rho$ and $k$ values allow reproduction of the total
radiative width of $^{172}Yb$ chosen for modelling. As it is seen
from Fig.~2, uncertainty of reproduction of the  spectra   $S_{if}$
using the data given in Fig.~1 is at least 5-10 times less than the error
of the experimental data [5] in spite of rather significant deviation of
$\rho$ and $k$ from the input values (as compared with estimations of
errors in [6]). Therefore, one can assume that the set of the $\rho$ and
$k$ values allowing reproduction of the experimental spectra $S_{if}$ is
considerably wider than that given by authors [6]. It should be noted that
the fluctuations of the $S$ values in Fig.~2 partially result from the
random fluctuations of the data in Fig.~1 owing to the use of the Monte-Carlo
method for their calculation.

The authors [5,6] point out that without the use of the experimental level
densities at two excitation energies, their experimental spectra can be
reproduced with infinite number of $\rho$ and $k=f A^{2/3}$ even if the
number of points in experimental spectra is ten times higher then number of
desired parameters. Hence, reliability of the physical conclusions in [6]
directly depends on systematical uncertainty of two experimental values $\rho$
used in [5,6]. It should be noted that level density for the considered
$^{172}Yb$ in vicinity of $B_n$ was determined experimentally only for
neutron resonances with $J^{\pi}=0^{-}$ and $1^{-}$. Reaction
$(^{3}He,\alpha)$ used in [5,6] excites (as it is pointed by authors)
mainly levels with spin 2 to 6. All conclusions of authors, however,
directly depends on accuracy in determination of level density in vicinity of
 $B_n$ just for the spin interval and parity of levels excited in the reaction.
But now it is impossible to evaluate the well known effect of rotational
enhancement  of level density in deformed nuclei on the results [6].
According [9], its value amounts to 10 for the complete level density.
Assuming that the parameters of rotational bands built on the excited states
$J<6$ of this nucleus insufficiently differ from that of rotational bands of
known low-lying states, and they do not depend on the parity of head level
one can estimate that rotation of the nucleus increases density of levels
$J=2-6$ by a factor of at least 2.5 as compared with the known model spin
dependencies.

An additional uncertainty can be also introduced in the results of this
analysis and if level density of even-even deformed nucleus at
$E_{ex}\approx 3-4$ MeV is less than that predicted by the simplest
extrapolations of $\rho$. This possibility follows from the
results [10] of extrapolation of the number of their cascade intermediate
levels to the zero detection threshold of the experiment.

Hence, in spite this uncertainty in level density at $\approx 8$ MeV, all
the data in Fig.~1 are mathematically equivalent (including level density in
the vicinity of $B_n$). And rejection from ``superfluous" solutions is
possible only at the presence of the complementary physics information.
Therefore, the question arises about the obtaining more reliable information
on the desired parameters of the cascade $\gamma$-decay process.

\section{Conclusion}
Correct estimation of systematical errors of the data like those given in
[5,6] first of all requires to take into account both sharp increase of
interval of values for $\rho$ and $k$, which allow reproduction of the
experimental spectra with the same precision due to strong their correlation,
and possible change in the extrapolated level density in given spin and
parity intervals due to known rotational (and, probably, vibrational) effect.

This work was supported by RFBR Grant $N^o$ 99-02-17863\\

1. R.Jacqmin, M.Salvatores, P.Bioux, D.Hittner,
International Conference on nuclear \\
\hspace*{14pt}data for science and technology.
Oct. 7-12, 2001, Japon, Abstracts, p. 7-O-2.\\
2. V.M. Maslov, Phys.Atomic Nuclei 63, 161 (2000)\\
3. E.V.Vasilieva, A.M.Sukhovoj, V.A.Khitrov,
Physics of Atomic Nuclei {\bf 64(2)} (2001) 153\\
\hspace*{14pt}V.A.Khitrov, A.M.Sukhovoj,
{\em Phys. of Atomic Nuclei\/} {\bf 64(7)} (2001) 1271\\
4. G.A. Bartholomew et al.,
{\em Advances in nuclear physics\/} {\bf 7} (1973) 229\\
5. A. Schiller et al., Nucl. Instrum. Methods Phys. Res. {\bf A447} (2000) 498\\
6. A.Voinov,  M. Guttormsen, E. Melby, J. Rekstad, A. Schiller,
 S. Siem, \\
\hspace*{14pt}Phys.\ Rev.\ C \bf 63(4)\rm,044313-1 (2001)
044309 (2001)\\
7. L.Aleksandrov, JINR Communications  P5-7259 (1973)\\
8. T.W.Anderson, An introduction to multivariate statistical analysis,
New York, Johm 
\\
\hspace*{14pt}Willey and Sons, Inc, London, Chapman and Hall, Limited, 1957.\\
9. "Handbook for calculation of nuclear reaction
data", IAEA-TECDOC-1034, 1998, \\
\hspace*{14pt}Vienna\\
10. A.M. Sukhovoj and V.A. Khitrov, Physics of Atomic Nuclei {\bf 62(1)},
(1999) 19\\


\newpage
\begin{figure}
[htbp]
\begin{center}
\leavevmode
\epsfxsize=15cm
\epsfbox{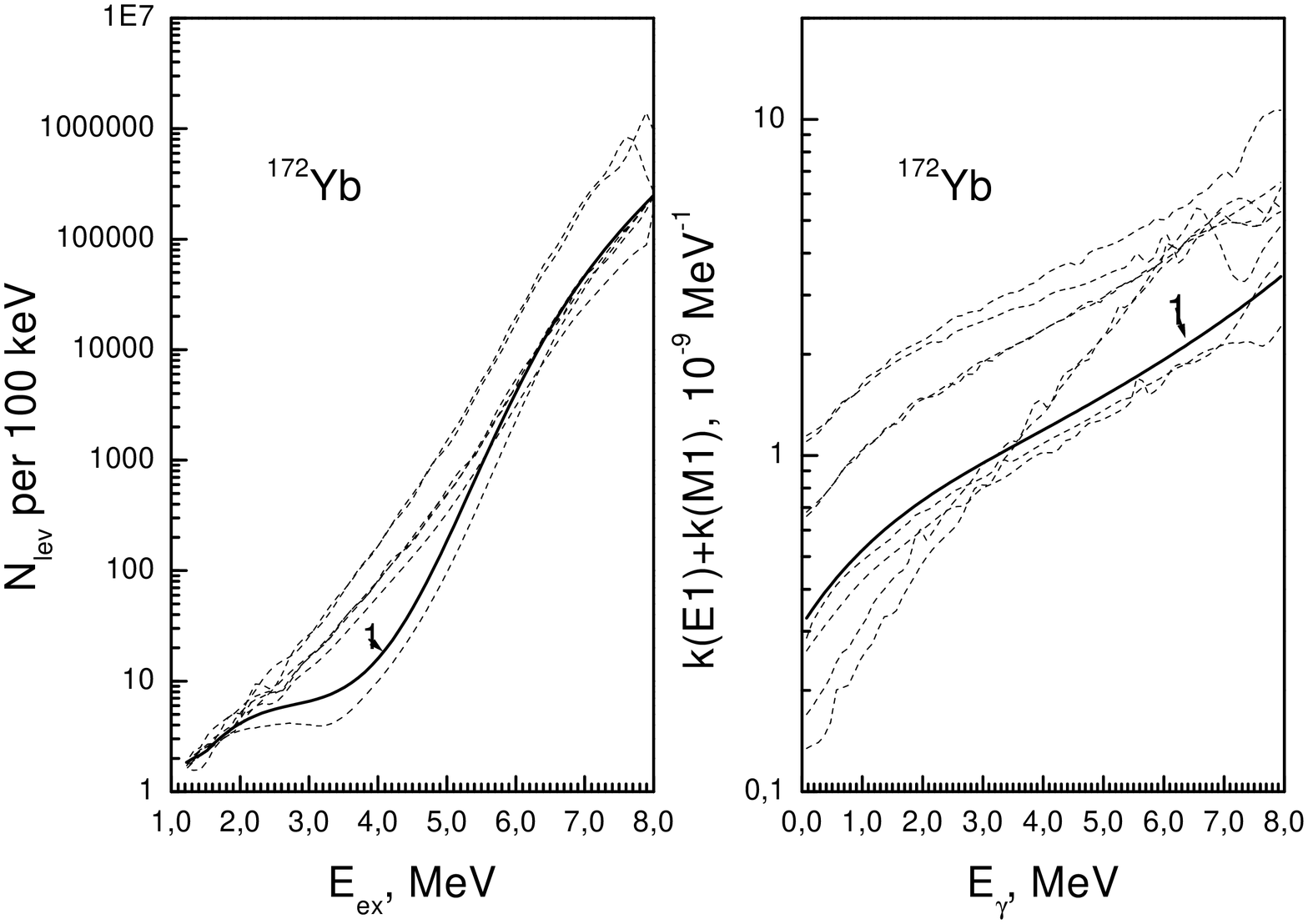}
\end{center}
\hspace{-0.8cm}
\vspace{1cm}

{\bf Fig.~1.}~Some random functional dependencies  of level density in
$^{172}Yb$ with $J=2\div 6$ (left panel) and radiative strength functions
(right panel) which allow reproduction the spectra of primary transitions
in Fig.~2. Curve 1 represents the desired values of  $\rho$ and $k$.
\end{figure}

\newpage

\begin{figure}[htbp]
\begin{center}
\leavevmode
\epsfxsize=15cm
\epsfbox{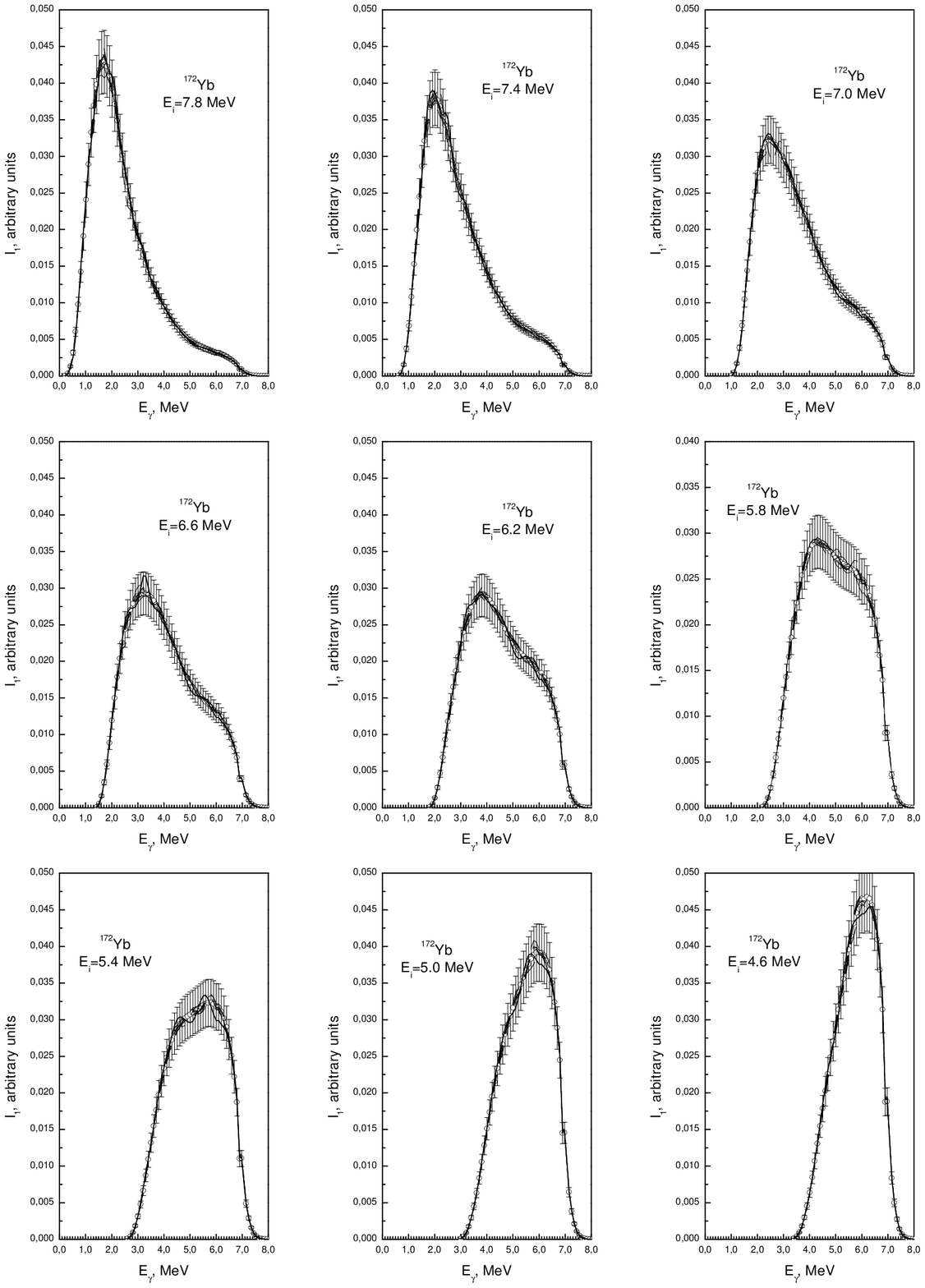}
\end{center}
\hspace{-0.8cm}

\vspace{-3cm}

{
\bf Fig.~2.}~Calculated spectra of the primary transitions for 
the energy $E_i$, corresponding to that given in [5]. 
Points with the error bars represent calculation
for curves ``1" in Fig.~1 with the 10\% interval of their deviations.
All spectra are shifted right to value $B_n-E_i$.
\end{figure}
\end{document}